\documentstyle[psfig]{l-aa}

\begin{document}

   \thesaurus{06         
              (03.11.1;  
               16.06.1;  
               19.06.1;  
               19.37.1;  
               19.53.1;  
               19.63.1)} 
   \title{Near-infrared photometry of the intermediate age open clusters
IC~166 and NGC~7789\thanks{Based on observations taken at TIRGO}}

   \author{Antonella Vallenari 
          \inst{1}, Giovanni Carraro\inst{2}, and
          Andrea Richichi\inst{3}
          }

   \offprints{Antonella Vallenari ({\tt vallenari@pd.astro.it})}

   \institute{Padova Astronomical Observatory, vicolo Osservatorio 
          5, I-35122, Padova,
	Italy
        \and
        Department of Astronomy, Padova University,
	vicolo dell'Osservatorio 5, I-35122, Padova, Italy
        \and
        Arcetri Astrophysical Observatory, Largo E. Fermi
        I-50110, Firenze, Italy\\
        e-mail: {\tt vallenari,carraro\char64pd.astro.it,
richichi\char64fi.astro.it}
             }

   \date{Received .....; accepted ......}

   \maketitle

   \markboth{Vallenari et al}{Infrared photometry}

   \begin{abstract}

We present and discuss new photometric data obtained with an IR camera
in the $J$ and $K$ pass-bands for the intermediate age open clusters
IC~166 (936 stars in total) and NGC~7789 (1030 stars in total). IC~166
is a poorly studied open cluster for which no IR data was available
previously, while NGC~7789 is a well studied open cluster.\\ We show
that IC~166 is of intermediate age (about 1.0~Gyr), with a reddening
E(V-K) $\approx 1.3$ (corresponding to E(B-V)~=~0.50) and a true distance modulus
$(m-M)_o~=~13.25$. These values are significantly different from
previous determinations.\\ NGC~7789 is found to be a 1.4 Gyr open
cluster. The metallicity derived with the method developed by Tiede et
al (1997) is found to be closer to the spectroscopic estimate than
previous photometric studies.

   \end{abstract}

%

\section{Introduction}

The Galactic Disk intermediate age open clusters are those clusters
whose ages range between the Hyades and IC~4651 (Carraro et al 1999a).
They are fundamental templates to study the internal structure of Main
Sequence (MS) stars with mass between $1.0$ and $2.0 M_{\odot}$. In
particular, they can be used to check the importance and the amount of
the core overshooting during the H-burning phase (Carraro et al 1993,
Rosvick \& Vandenberg 1998).

Usually the comparison between observational data and theoretical
models is done by means of the Color Magnitude Diagrams (CMDs) and the
dating operation of the stellar ensembles by means of isochrone
fitting, luminosity functions and star counts. For stellar clusters,
obtaining reasonable age estimates requires knowledge of cluster
reddening, distance and metallicity. IR photometry is particularly
useful to obtain cluster reddening when combined with optical
photometry. Additionally, it is possible to get an estimate of the
cluster metallicity whenever a Red Giant Branch (RGB) is visible
(Tiede et al. 1997). 

This study is a part of a general project of observations of galactic
open clusters in the infrared (Vallenari et al. 1999, Carraro et al.
1999b). In this paper we present IR camera photometry in the $K$ and
$J$ pass-bands for two intermediate age open clusters, IC~166 and
NGC~7789.  IC~166 is a poorly studied open cluster for which no
previous IR data exist, while NGC~7789 is a rather well-studied open
cluster, for which up to now IR data were available
only for a handful of
stars. Table~1 summarizes some general properties of the two target
clusters. Diameters are taken from the Lyng{\aa} (1987) catalogue of
open clusters.

The plan of the paper is as follows. In Section~2 we describe the data
acquisition and reduction; Section~3 is devoted to the analysis of the
data for IC~166, while Section~4 is dedicated to NGC~7789.  Section~5
draws our conclusions.

\begin{table}
\tabcolsep 0.10truecm
\caption{Basic parameters of the studied clusters. Diameters
are taken from Lyng{\aa} (1987) catalogue.}
\begin{tabular}{lccccc} 
\hline
\hline
\multicolumn{1}{c}{Cluster} &
\multicolumn{1}{c}{$\alpha_{2000.0}$} &
\multicolumn{1}{c}{$\delta_{2000.0}$} &
\multicolumn{1}{c}{$l$} &
\multicolumn{1}{c}{$b$} &
\multicolumn{1}{c}{Diameter}\\
 &$hh~mm$& $^{o}$ &$^{o}$ &$^{o}$ &($\prime$) \\
\hline
IC~166         & 01~52~30.8 & +61~35~47.5 & 130.08 & -0.19 &  7\\
NGC~7789       & 23~57~02.0 & +54~46~42.1 & 115.49 & -5.36 & 25\\
\hline\hline
\end{tabular}
\end{table}

\section{Observations and data reduction}

$J$ (1.2 $\mu$m) and $K$ (2.2 $\mu$m) photometry of the two clusters was
obtained with 1.5m Gornergrat Infrared Telescope (TIRGO) equipped with
the Arcetri Near Infrared Camera (ARNICA) in October 1997.  ARNICA
relies on a NICMOS3 256$\times$ 256 pixels array (gain=20 e$^-/$ADU,
read-out noise=50 e$^-$ angular scale=1$\arcsec/$pixel, and $4 \times
4 $arcmin field of view).  Through each filter 4 partially overlapping
images of each cluster were obtained, covering a total field of view
of about 8 $\times 8$ arcmin, in short exposures to avoid sky
saturation.  Details of the observations are given in Table~2.  The
night was photometric with a seeing of
1$\arcsec$-1.5$\arcsec$. Figs.~1 and 2 present the mosaics of the 4
frames obtained per cluster in $K$ passband.

The data were reduced subtracting from each image a linear combination
of the corresponding skies and dividing the results by the
flat-field. We make use of the Arnica package (Hunt et al. 1994) in
IRAF and Daophot II.  The conversion of the instrumental magnitude $j$
and $k$ to the standard $J$, $K$ was made using stellar fields of standard
stars taken from Hunt et al. (1998) list.  About 10 standard stars per
night have been used.  The relations in usage per 1 sec exposure time
are for IC~166:
 
\begin{equation}
J  = j+19.51 + k_J \times 1.04\\
\end{equation}

\begin{equation}
K  = k+18.94 + k_K \times 1.06\\
\end{equation}

\noindent 
and for NGC~7789:

\begin{equation}
J  = j+19.51 + k_J \times 1.04\\
\end{equation}

\begin{equation}
K  = k+18.94 + k_K \times 1.03\\
\end{equation}

\noindent
where $k_J$ and $k_K$ (the extinction coefficients, in magnitudes per airmass) 
are 0.25 and 0.10, respectively.

\begin{figure}
\centerline{\psfig{file=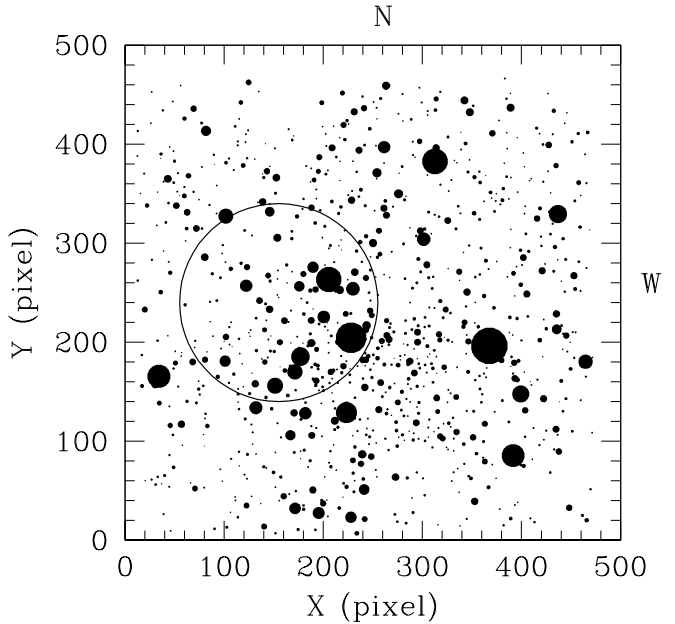,width=9cm,height=9cm}}
\caption{A mosaic of the four CCD frames in K band covering the studied region of IC~166. North 
is on the top, east on the left. The circle
defines the region of the cluster observed by Burkhead (1969).}
\end{figure}

\begin{figure}
\centerline{\psfig{file=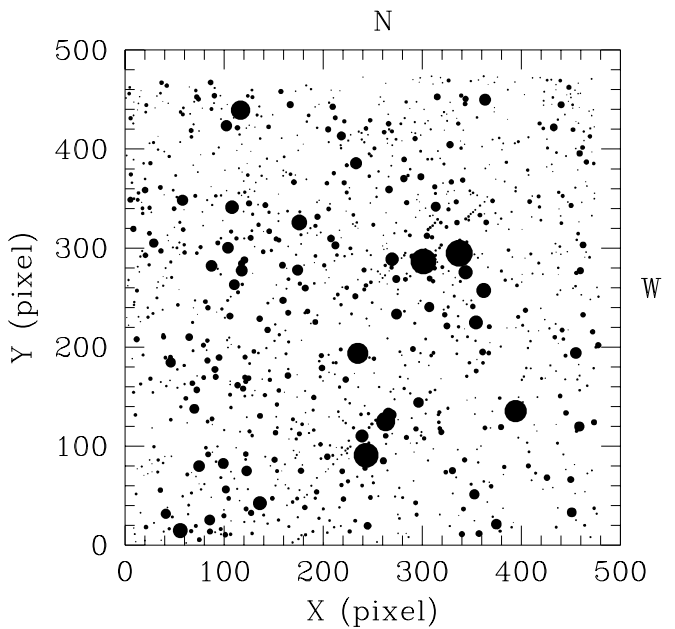,width=9cm,height=9cm}}
\caption{A mosaic of the four CCD frames in K band covering the studied region of NGC~7789. 
North is on the top, east on the left.}
\end{figure}

\begin{figure*}
\centerline{\psfig{file=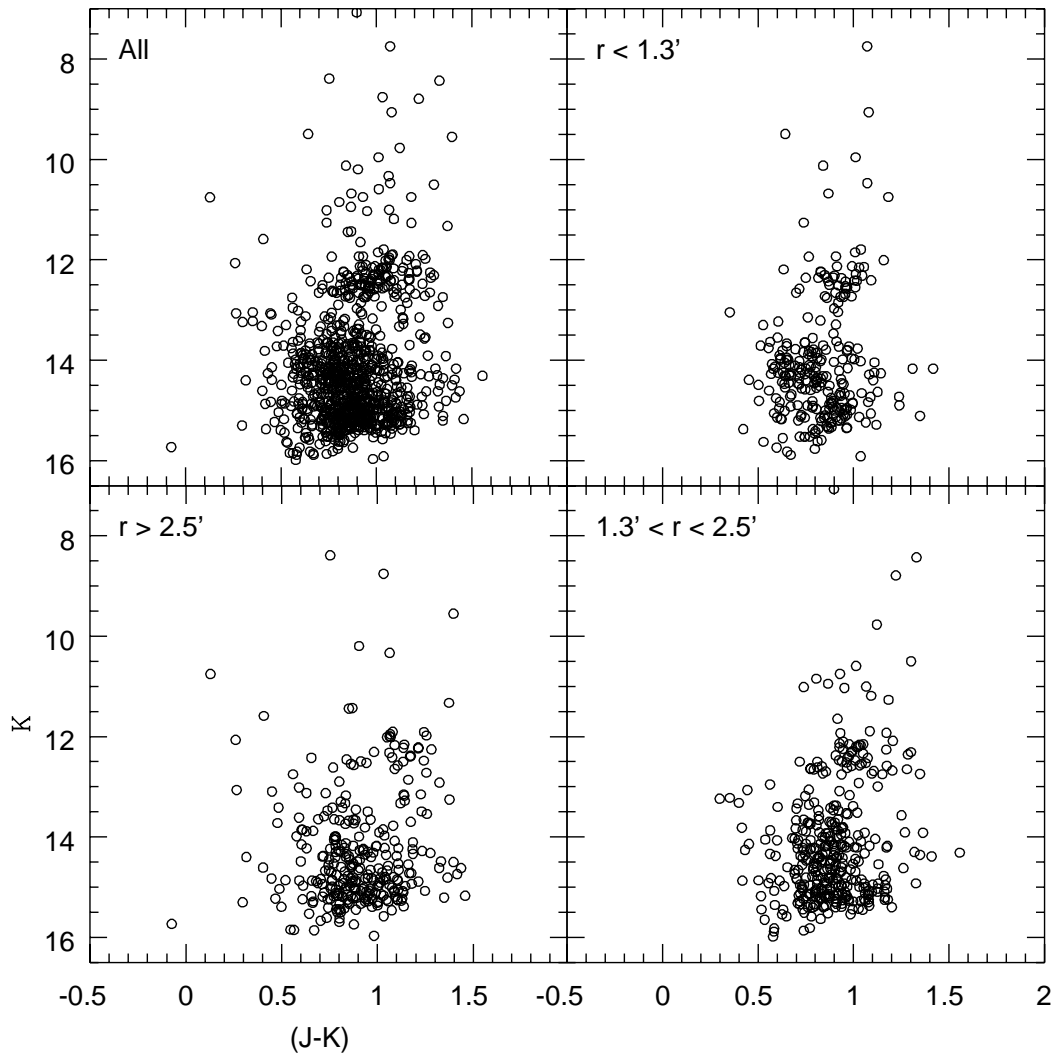,width=16cm,height=16cm}}
\caption{CMD for IC~166 in different areas of the covered region.}
\end{figure*}

\begin{table*}
\caption[ ]{ Observation Log-Book. The coordinates listed below refer to
the center of the mosaic.}
\tabcolsep 0.8truecm
\begin{tabular}{c|c|c|c|c|c}
\hline
\hline
Cluster               &$\alpha$    &$\delta$  & Date& 
\multicolumn{2}{c}
{Exposure Times (sec)} \\ 
                           &(2000)      &(2000) &    &J&K\\
\hline
IC 166    &01 52 21.6  & 61 52 12 & Oct, 24, 1997& 700 & 920 \\
NGC 7789  &23 57 10.9 & 56 45 05 & Oct, 24, 1997& 636 & 920 \\
\hline
\hline
\end{tabular}
\end{table*}

\noindent 
The standard deviation of the zero points are  0.03  mag for the J 
and 0.04 for the K magnitude. This error is only due to the linear
interpolation of the standard stars.  However the calibration
uncertainty is dominated by the error due to the correction from
aperture photometry to PSF fitting magnitude.  
Taking this into account,
we estimate that the total calibration error is about 0.1 mag
both in $J$ and in $K$ passbands. The photometric errors,
as produced by DAHPHOT are 0.02, 0.05 and 0.1 at $J$ equal to
8, 12 and 16 mag, respectively, and slightly lower in the $K$ band.
Accordingly, the maximum error in the color amounts to about 0.25.\\
Fitting photometry was a natural
choice, due to the number of stars to be measured and the
concentration of stars (crowding) in some cluster regions, where PSF
wings overlap.  The standard stars used for the calibration do not
cover the entire colour range of the data, because of the lack of
stars redder than $(J-K) \sim 0.8$. From our data, no colour term is
found for K mag, whereas we cannot exclude it for the J magnitude.

The data tables are available upon request from the authors.

\section{IC~166}

IC~166 (C0149+615, OCL~334, Trumpler class II 1 r) is a faint, distant
and possibly old cluster (King 1964).  The only photometric study was
done by Burkhead (1969), who obtained BV photographic photometry for
about 200 stars in the central region of the cluster, covering a
circular area of $2.45 ^{\prime}$ (see Fig.~1) and reaching
$V~=~19.0$.  He reported also UBV photo-electric photometry for 20
stars out of the cluster region.  The derived CMD shows a wide MS,
with the Turn-off-point (TO) located at $V \approx 17.0$ and $(B-V)
\approx 1.1$; the Herzsprung gap and a conspicuous clump of red stars
are also seen.  There is no evidence for RGB stars, which implies that
the cluster is intermediate in age between the Hyades and NGC~7789
(Carraro et al. 1999).  Burkhead (1969) could not determine the
reddening due to the lack of photometry in U band for cluster members.
Assuming E(B-V) $\approx$ 0.80, he found a distance modulus $(m-M)_V
\approx$ 12.6, and a distance of 3.3~kpc from the Sun.  Finally,
inspecting the Palomar Sky Atlas, he estimated a diameter of about 5
arcmin.

\begin{figure}
\centerline{\psfig{file=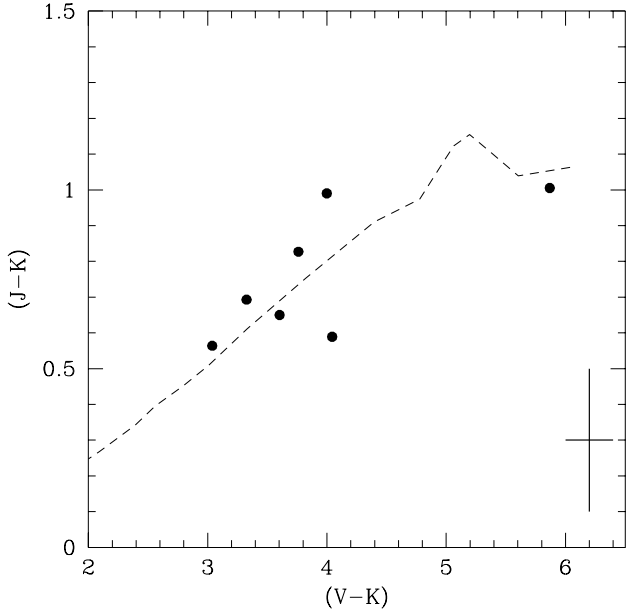,width=9cm,height=9cm}}
\caption{Two colors diagram for MS stars in common between our study and Burkhead(1969). Dashed 
line is a ZAMS from Bertelli et al. (1994) shifted by 
$E_{(V-K)}~=~1.35$ and $E_{(J-K)}$~=~0.25. The cross in the right bottom corner shows the 
photometric error bar indicating the maximum error in colors for the data points.}
\end{figure}

\begin{figure}
\centerline{\psfig{file=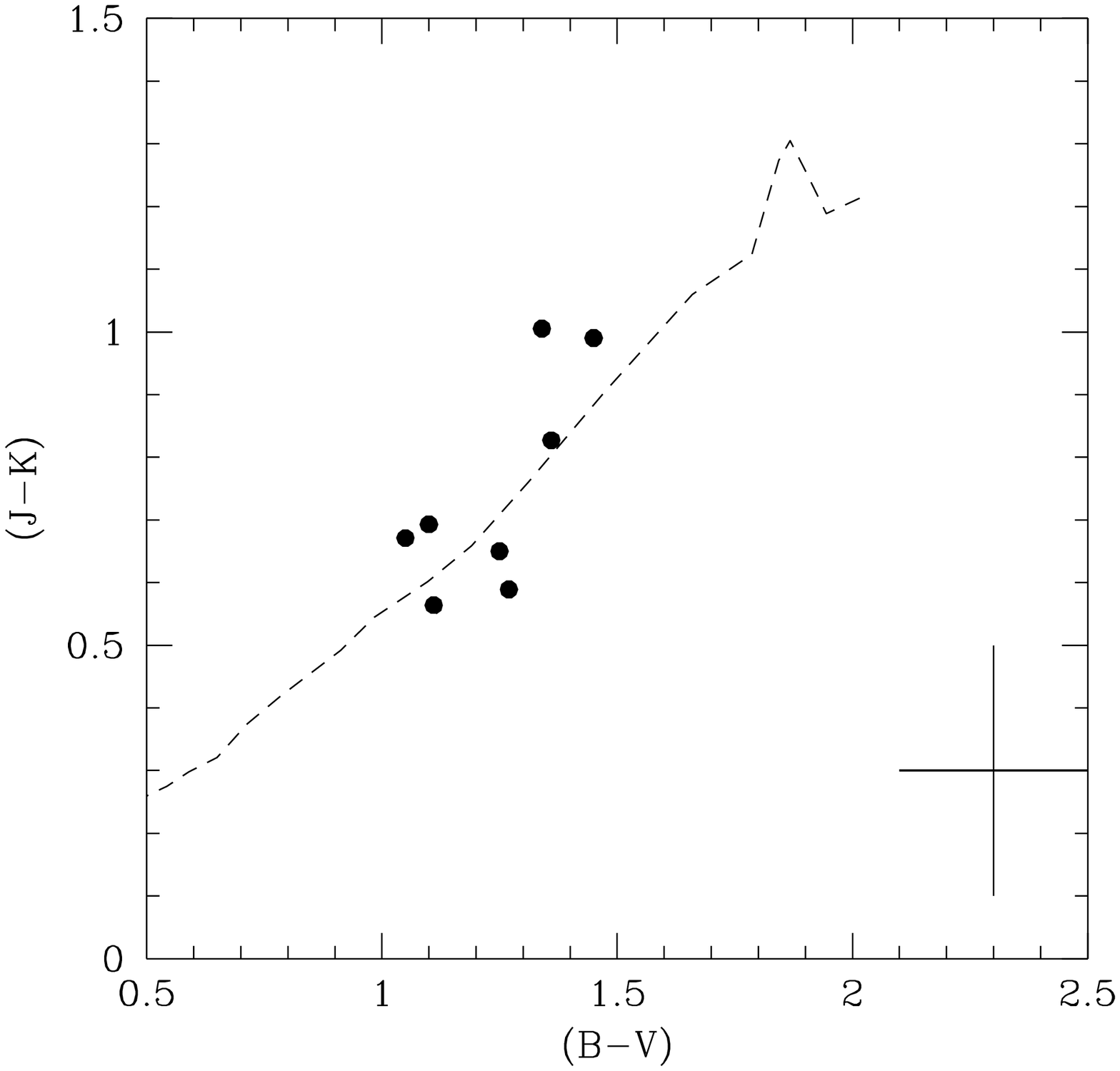,width=9cm,height=9cm}}
\caption{Two colors diagram for MS stars in common between
our study and Burkhead(1969). Dashed line is a ZAMS from
Bertelli et al. (1994) shifted by 
$E_{(B-V)}~=~0.50$ and $E_{(J-K)}$~=~0.25. The cross in the 
right bottom corner shows the photometric error bar indicating the maximum error in color
for the data points.}
\end{figure}

\begin{figure}
\centerline{\psfig{file=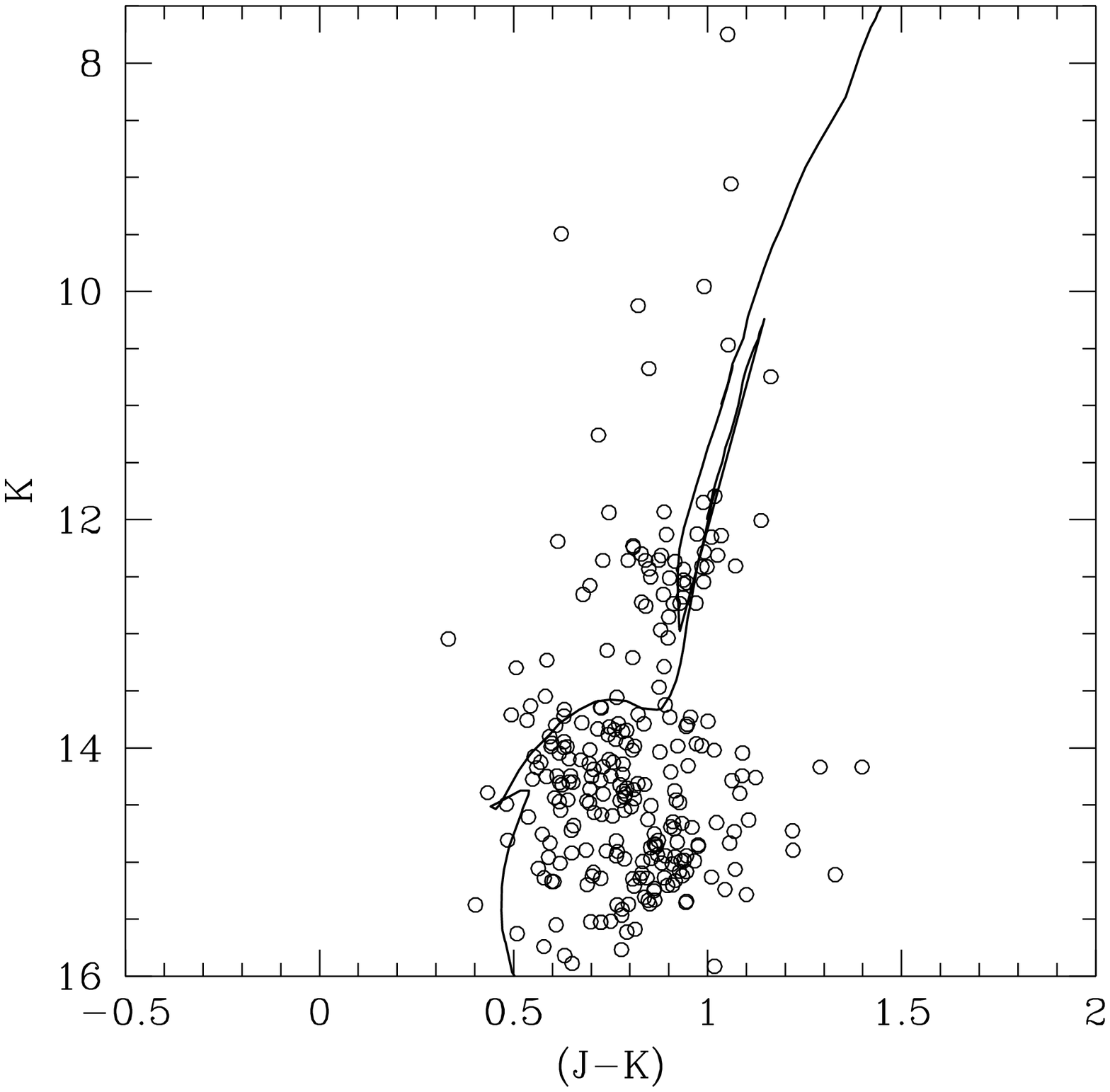,width=9cm,height=9cm}}
\caption{The CMD for all the stars in the central region of
IC~166. Superimposed is a $Z~=~0.009$ isochrone for an age of
$1~Gyr$. See the text for any detail.}
\end{figure}

\subsection{The Color-Magnitude Diagram}

The CMD of IC~166 in the plane $K$ vs $(J-K)$ is shown in Fig.~3.  The
upper left panel shows the CMD for all the detected stars, the upper
right for the stars lying in the same region as Burkhead (1969)
photometry (see also Fig.~1). The lower left panel shows the stars
outside $2.5 ^{\prime}$ from the cluster center, and finally the lower
right shows the CMD in the region between about $1.3^{\prime}$ and
$2.5^{\prime}$ from the cluster center.  The cluster center has been
fixed on the top of the star \# 1 in the Burkhead second quadrant.

We cover a region of $8^{\prime} \times 8^{\prime}$, whose center is
offset by $1^{\prime}$ with respect to the Burkhead (1969) photometry.
Nonetheless we cover almost all the cluster region.  Our CMD shows a
wide MS and a sparse RGB clump. The MS extension is the same as the
optical one from Burkhead (1969).  The contamination of foreground
stars is considerable in the upper left panel, whereas the cluster
clearly dominates in the upper and lower right panels.  A
representative contaminating disk population is better visible in the
lower left panel, outside the Burkhead (1969) radius.  This confirms
that the Burkhead (1969) radius estimate ($\approx 2-3$ arcmin) is
basically correct.

From the global CMD morphology we can confirm that IC~166 is an
intermediate age open cluster, as old as NGC~2477 (1.0 Gyr, Carraro \&
Chiosi 1994), but younger than NGC~7789 (1.6 Gyr, Gim et al. 1998).

\subsection{Reddening}

Useful information about the reddening of IC~166 can be derived by
combining optical and infrared photometry.  We found 30 stars in
common between our study and Burkhead (1969).  Singling out the MS
stars, we are left with 8 stars.  These are plotted in the plane (J-K)
vs (V-K) (see Fig.~4) and (J-K) vs (B-V) (see Fig.~5). Superimposed
are Zero Age MS for Z = 0.009 metallicity taken form Bertelli et
al. (1994). The fit in Fig.~4 has been obtained by shifting the ZAMS
with E(V-K) = 1.35 $\pm$ 0.50 and E(J-K) = 0.25 $\pm$ 0.10,
corresponding to a ratio E(V-K) / E(J-K) $\approx$ 5.4, close to the
value 5.3 reported by Cardelli et al. (1989). From Fig.~5 we get
E(B-V) = 0.50 $\pm$ 0.20 and E(J-K) = 0.25 $\pm$ 0.10.  This gives a
ratio E(J-K) / E(B-V) $\approx$ 0.50 which again is close to the value
0.52 from Cardelli et al. (1989).  These results, although consistent,
are to be taken as provisional, until a deeper optical and IR
photometry will be available.

\subsection{Distance and age}

Distance and age are inferred by fitting the CMD of IC~166 (see
Fig~.6) with theoretical isochrones (Bertelli et al. 1994). 

The metal content of IC~166 has been determined by Friel \& Janes
(1993) using moderate resolution spectroscopy of 4 giant stars. [Fe/H]
is -0.32 $\pm$ 0.20, which translates into the theoretical metal
abundance value Z = 0.009 (Bertelli et al. 1994).  Adopting the
reddening value derived above (E(J-K)~=~0.25), 
and the spectroscopic metallicity from
Friel \& Janes (1993), we obtain an acceptable fit with a 1.0 Gyr
isochrone. In fitting the isochrone to the CMD we have looked at the
TO magnitude and at the mean clump magnitude, since the shape of the
MS below the TO is not well defined. This implies that it is difficult
to give an error to the age until a deeper photometry is available.

The mean clump magnitude allows us to derive the apparent distance
modulus. We obtain $(m-M)_K \approx$ 14.00 $\pm$ 0.30, but the true
distance modulus turns out to be $(m-M)_{o,K} \approx$ 13.25, which
puts IC~166 4.5 $\pm$ 0.6 kpc from the Sun.  This value for the
distance is significantly larger than the Burkhead (1969) estimate. We
ascribe this result to the different value we find for the cluster
reddening.

\begin{figure}
\centerline{\psfig{file=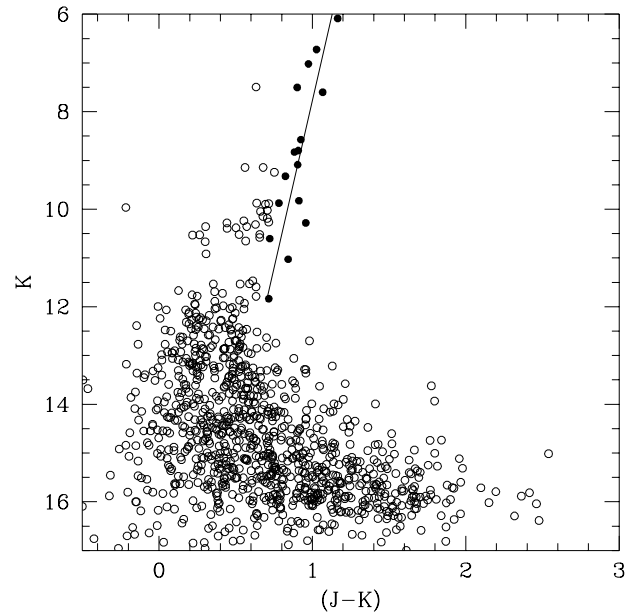,width=9cm,height=9cm}}
\caption{The CMD for all the stars in the region of
NGC~7789. The solid line represents a least squares fit
to the presumed RGB stars (filled circles).}
\end{figure}

\section{NGC~7789}

NGC~7789 is a populous intermediate age open cluster very well studied
in the past. The most recent photometry is from Gim et al. (1998), who
studied 15,000 stars within a radius of $\approx 18^{\prime}$ from the
cluster center, and which the reader is referred to for more detailed
informations on this important cluster. 

The fundamental parameters of this cluster have been determined
several times: NGC~7789 is about 1.6 Gyr old (Gim et al. 1998); the
available metallicity determinations range from -0.26 to -0.62, as
measured by the index [Fe/H] (Friel \& Janes 1993, Tiede et al. 1997);
reddening amounts to E(B-V) = 0.35, while true distance modulus is
$(m-M)_0~=~11.30$. Near IR photometry is reported by Manteiga et
al. (1991) for a sample of 14 presumed blue stragglers in the field of
NGC~7789 to test the binary hypothesis for these stars, and by Frogel
\& Elias (1988) for a sample of 10 bright red giants to study the mass
loss mechanism during the RGB climbing. We find 6 stars in common
with Frogel \& Elias (1988), namely the stars $\#193$, $\#304$,
$\#329$, $\#494$, $\#501$ and $\#669$, according to the K\"ustner
(1923) numbering. The photometric comparison of these stars provides a
good agreement, with the mean differences being:

\begin{equation}
K_{FE} - K_{VCR} = -0.04 \pm 0.02
\end{equation}

\begin{equation}
(J-K)_{FE} - (J-K)_{VCR} = 0.05 \pm 0.03
\end{equation}

\noindent
where the suffix $FE$ refers to Frogel \& Elias (1988), and $VCR$
to the present study. The errors reported are the standard deviations
around the mean values. Using this small sample we cannot outline
any clear trend with  colors or magnitudes.

\subsection{The Color Magnitude Diagram}

Our study covers a region within $8^{\prime}$ of the cluster center
(see Fig.~2).  We obtain photometry for about 1030 stars down to $K
\approx 17.0$. This allows us to present the first CMD of NGC~7789 in
the infrared. This is shown in Fig.~7.  The MS is quite wide, and
shows the TO at $K \approx 12.0$ and $(J-K) \approx 0.25$. A prominent
clump is situated at $K \approx 10.5$ and $0.4 < (J-K)< 0.7$. The RGB
is defined by the stars plotted by us with a different symbol (filled
circles). 

The overall morphology is typical of an intermediate age open cluster
(Carraro et al. 1999a).

\subsection{Metallicity}

The CMD diagram in the IR allows us to derive an independent estimate
of the cluster abundance by using the photometric method originally
developed by Kuchinski \& Frogel(1995) for metal rich globular
clusters, and then applied by Tiede et al. (1997) for a sample of
intermediate age open clusters.  This method correlates the slope of
the RGB, defined as $\Delta(J-K)/\Delta K$ with the cluster
metallicity, measured by the index [Fe/H]. For globulars the relation
reads:

\begin{equation}
[Fe/H] = -2.98 - 23.84 \times (GB slope).
\end{equation}

\noindent
Tiede et al. (1997) found that equation (7), when applied to open
clusters rather than globulars, tends to underestimate open cluster
metallicity. For instance, in the case of NGC~7789 they found [Fe/H] =
-0.62 by using equation (7), whereas the spectroscopic determination
is [Fe/H] = -0.26. For this reason they give a new calibration of the
relation, which provides new values for the coefficients when dealing
with different populations (globulars, open, or bulge clusters).  For
open clusters, the modified relation reads:

\begin{equation}
[Fe/H] = -1.639 - 14.243 \times (GB slope).
\end{equation}

\begin{figure}
\centerline{\psfig{file=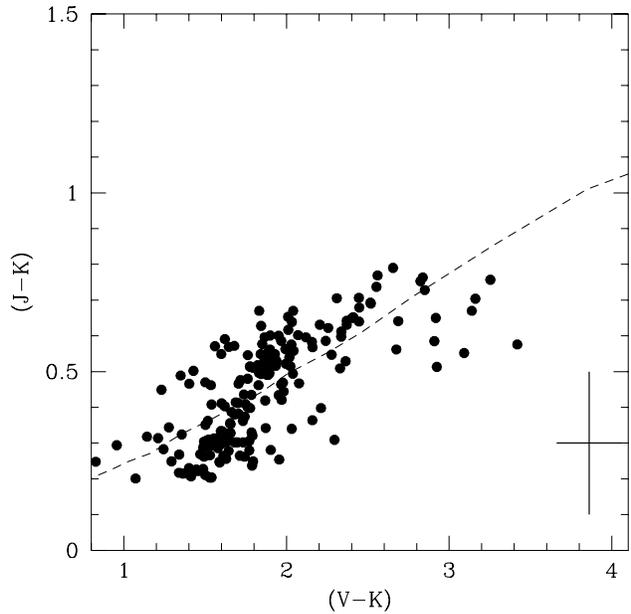,width=9cm,height=9cm}}
\caption{Two color diagram for the MS stars brighter than J =14
in common between our study and Gim et al(1998). Dashed line is a Z= 0.010 ZAMS taken from 
Bertelli et al. (1994) shifted by E(V-K) = 0.85 and E(J-K) = 0.15. 
The cross in the bottom right 
corner shows the photometric error bars indicating the maximum 
error in color for the data points.}
\end{figure}

\begin{figure}
\centerline{\psfig{file=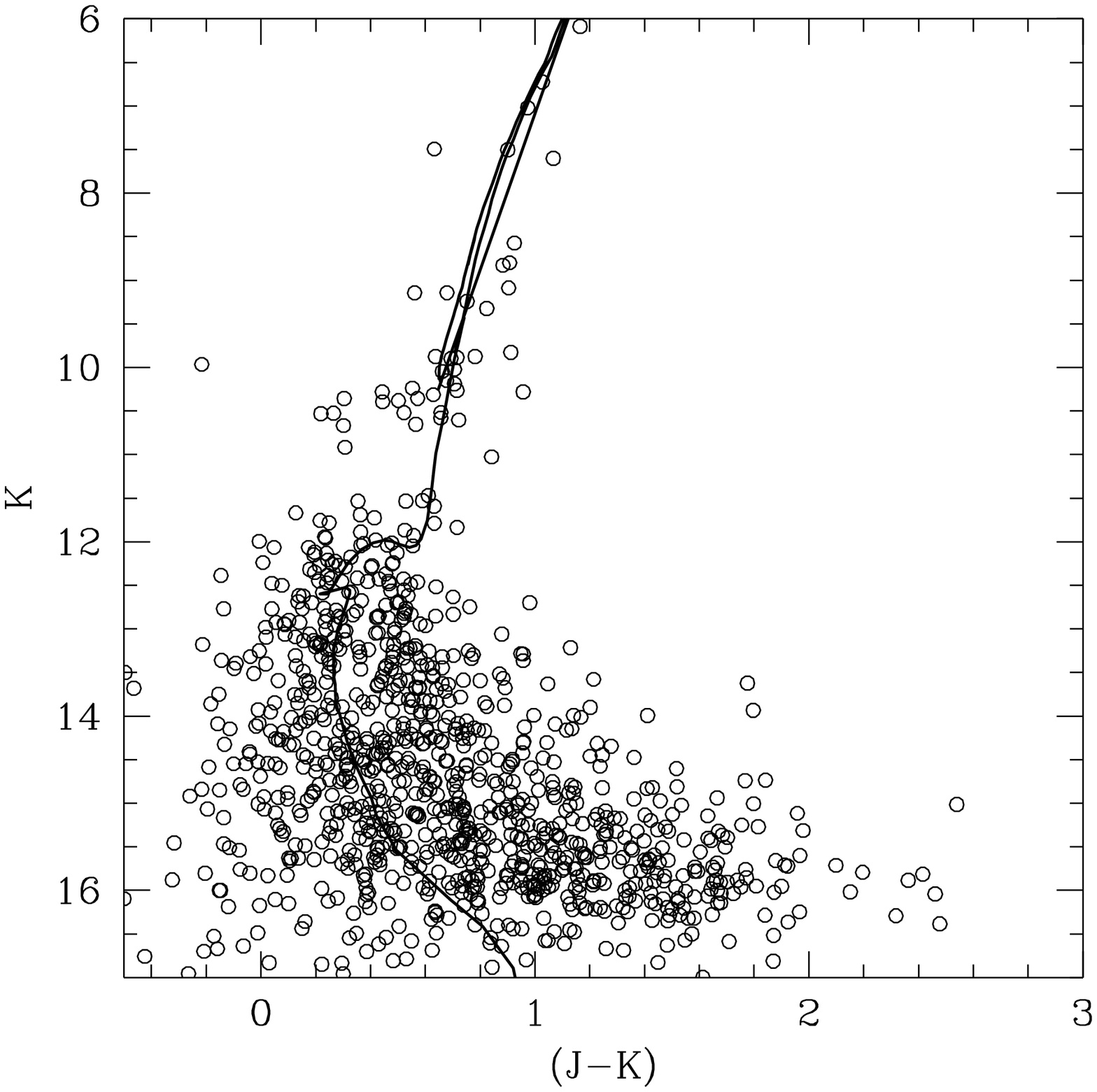,width=9cm,height=9cm}}
\caption{The CMD for all the stars in the region of
NGC~7789. Superimposed is a $Z~=~0.010$ isochrone for an age
of $1.4~Gyr$. See text for any detail.}
\end{figure}

\begin{figure}
\centerline{\psfig{file=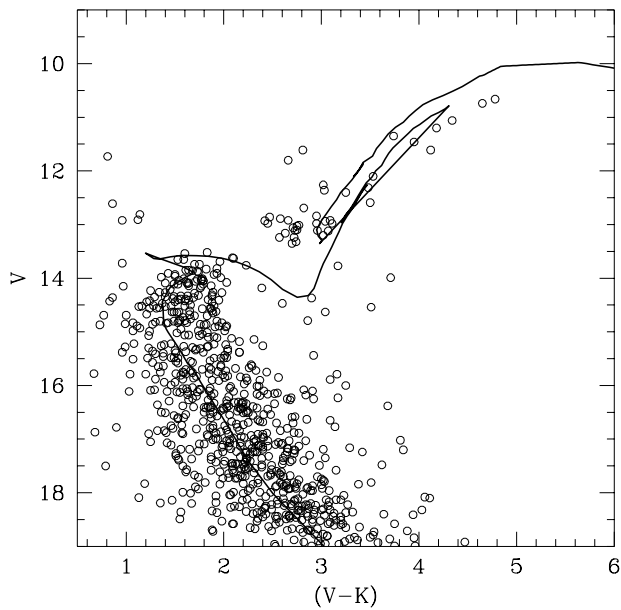,width=9cm,height=9cm}}
\caption{The CMD for all the stars in the region of
NGC~7789. Superimposed is a $Z~=~0.010$ isochrone for an age
of $1.4~Gyr$. See text for any detail.}
\end{figure}

\noindent To find the RGB slope we performed a least squares fit to
the RGB stars as indicated in Fig.~7. This implies an RGB slope
$\Delta(J-K) / \Delta K$ of -0.097 $\pm$ 0.007.  By using the relation
(8), we obtain [Fe/H] = -0.25 $\pm$ 0.11.  The reported error has to
be considered as an optimistic estimate, since it does not take into
account the uncertainties in the coefficients of eq.~8, and the
sensitivity of the RGB slope to the method adopted for its
computation. However the value we find implies a metal content close
the spectroscopic estimate (-0.26 $\pm$ 0.06, Friel \& Janes 1993).

\subsection{Reddening}

As for IC~166, we combine together optical and infrared photometry.
We found 980 stars in common between our study and the optical
photometry of Gim et al. (1998). Out of these we consider MS the stars
brighter than $J$ = 14 and in the color interval $0.2 <(J-K)< 0.8$, to
limit disk stars contamination.  These MS stars are plotted in
Fig.~8. Although the scatter is large, a reasonable fit can be
obtained shifting a Z = 0.010 ZAMS taken from Bertelli et al. (1994)
by E(J-K) = 0.15 $\pm$ 0.06 and E(V-K) = 0.85 $\pm$ 0.20.  This way
the ratio $E_{(V-K)} / E_{(J-K)}$ comes out to be $\approx 5.6$, close
to the value $5.3$ reported by Cardelli et al. (1989). Adopting the
value $0.52$ (Cardelli et al. 1989) for the ratio $E_{(J-K)} /
E_{(B-V)}$ we obtain E(B-V) $\approx$ 0.30.

\subsection{Distance and Age}

Assuming the metallicity to be $Z~=~0.010$, we have performed a fit to
the CMD (see Fig.~9) with a 1.4 Gyr isochrone.  The best solution has
been obtained by adjusting the isochrone with a color excess E(J-K) =
0.15 and an apparent distance modulus $(m-M)_K~=~11.70$.  This implies
a true distance modulus $(m-M)_o~=~11.25$, close to the accepted
estimate, and a color excess E(B-V) = 0.29, in agreement with the
solution derived from the two colors diagram, but marginally smaller
than the most accepted estimate. 

The fundamental parameters derived above are supported by the fit we
performed in the $V$ vs $(V-K)$ plane (see Fig.~10), where the
superimposed 1.4 Gyr isochrone has been shifted by E(V-K) = 0.75 and
$(m-M)_V~=~13.20$. We point out that looking at Fig.~9, it is not
possible to reproduce the color of the RGB.

\section{Discussion and Conclusions}

We have presented and discussed new IR camera data for two
intermediate age open clusters, IC~166 and NGC~7789. IC~166 was poorly
studied before, while NGC~7789 is a very well studied cluster.  Our
results can be summarized as follows:

\begin{description}
\item{$\bullet$} IC~166 is a faint and distant intermediate age open cluster about 1 Gyr old;
\item{$\bullet$} for the first time we are able to determine the reddening 
and distance modulus of this cluster, which is located 4.5 kpc from the Sun;  much deeper 
photometry is required to obtain a better comparison with stellar models.
\item{$\bullet$} NGC~7789 is shown to be 1.4 Gyr old; we find estimates for the cluster 
parameters consistent with values in the literature;
\item{$\bullet$} we obtain an independent photometric metallicity estimate which is close to the 
spectroscopic one.
\end{description}

\begin{acknowledgements}
We acknowledge the anonymous referee for the detailed report
on the first version of the paper which helped to improve
the presentation of this article.
This research has been sponsored by the Italian Ministry
of University and Research, and by the Italian Space Agency.
\end{acknowledgements}

{}

\end{document}